\documentclass[a4paper,conference]{IEEEtran}

\usepackage{fancyhdr}
\usepackage{xparse}

\usepackage[ruled,vlined,linesnumbered]{algorithm2e}

\usepackage{amsmath}
\usepackage{amstext}
\usepackage{amssymb}
\usepackage{graphicx}
\usepackage{hyphenat}
 \usepackage{color}
 \usepackage{cite}
 
\usepackage[T1]{fontenc}
\usepackage[utf8]{inputenc}
\usepackage{authblk}

\newcommand{\abs}[1]{\ensuremath{\left| #1 \right|}}

\NewDocumentCommand\col{g}{%
  \IfNoValueTF{#1}{\ensuremath{\mathrm{vec}}}{\ensuremath{\mathrm{vec}}\of{#1}}%
}

\NewDocumentCommand\of{og}{%
  \IfNoValueTF{#1}%
    { \IfNoValueTF{#2}{}{\!\({#2}\)} }%
    { \IfNoValueTF{#2}{\!\[{#1}\]}{\!\{{#2}\}} }%
}

\DeclareMathOperator{\Diff}{\ipaclap{D}{\raisebox{.204em}{\textpalhook}\kern.44em}\kern-.1em}
\NewDocumentCommand\diff{g}{%
  \IfNoValueTF{#1}
  {\text{\texthtd}}
  {\text{\texthtd}\of{#1}}%
}

\RenewDocumentCommand\ln{g}{%
  \IfNoValueTF{#1}{\mathrm{ln\ }}{\mathrm{ln}\of{#1}}%
}

\NewDocumentCommand\Real{og}{%
  \IfNoValueTF{#1}%
    { \IfNoValueTF{#2}{\mathcal{R}\!\!\mathpzc{e}}{\mathcal{R}\!\!\mathpzc{e}\!\{{#2}\}} }%
    { \IfNoValueTF{#2}{\mathcal{R}\!\!\mathpzc{e}\!\[{#1}\]}{\mathcal{R}\!\!\mathpzc{e}\!\({#2}\)} }%
}
\NewDocumentCommand\Imag{og}{%
  \IfNoValueTF{#1}%
    { \IfNoValueTF{#2}{\mathcal{I}\!\!\mathpzc{m}}{\mathcal{I}\!\!\mathpzc{m}\!\{{#2}\}} }%
    { \IfNoValueTF{#2}{\mathcal{I}\!\!\mathpzc{m}\!\[{#1}\]}{\mathcal{I}\!\!\mathpzc{m}\!\({#2}\)} }%
}

\RenewDocumentCommand\cos{g}{%
  \IfNoValueTF{#1}{\mathrm{cos}}{\mathrm{cos}\of{#1}}%
}
\RenewDocumentCommand\sin{g}{%
  \IfNoValueTF{#1}{\mathrm{sin}}{\mathrm{sin}\of{#1}}%
}
\RenewDocumentCommand\tan{g}{%
  \IfNoValueTF{#1}{\mathrm{tan}}{\mathrm{tan}\of{#1}}%
}
\RenewDocumentCommand\arccos{g}{%
  \IfNoValueTF{#1}{\mathrm{arccos}}{\mathrm{arccos}\of{#1}}%
}
\RenewDocumentCommand\arcsin{g}{%
  \IfNoValueTF{#1}{\mathrm{arcsin}}{\mathrm{arcsin}\of{#1}}%
}
\RenewDocumentCommand\arctan{g}{%
  \IfNoValueTF{#1}{\mathrm{arctan}}{\mathrm{arctan}\of{#1}}%
}
\RenewDocumentCommand\cot{g}{%
  \IfNoValueTF{#1}{\mathrm{cot}}{\mathrm{cot}\of{#1}}%
}

\newcommand{\floor}[1]{\ensuremath{\left\lfloor #1 \right\rfloor}}

\NewDocumentCommand\tr{g}{%
  \IfNoValueTF{#1}{\mathrm{tr}}{\mathrm{tr}\of{#1}}%
}

\NewDocumentCommand\diag{og}{%
  \IfNoValueTF{#1}%
    { \IfNoValueTF{#2}{\ensuremath{\mathrm{diag}}}{\ensuremath{\mathrm{diag}\of{#2}}} }%
    { \IfNoValueTF{#2}{\ensuremath{\mathrm{diag}\of[#1]}}{\ensuremath{\mathrm{diag}\of[]{#2}}} }%
}

\RenewDocumentCommand\exp{g}{%
  \IfNoValueTF{#1}{\ensuremath{\mathrm{exp}}}{\ensuremath{\mathrm{exp}}\of{#1}}%
}

\newcommand{\E}[2][]{\Operator[#1]{E}{#2}}

\newcommand{\R}[2][]{\Operator[#1]{R}{#2}}

\NewDocumentCommand\C{g}{%
  \IfNoValueTF{#1}{\mathrm{Cov}}{\mathrm{Cov}\of{#1}}%
}

\renewcommand{\(}{\ensuremath{\left(}}
\renewcommand{\)}{\ensuremath{\right)}}
\renewcommand{\[}{\ensuremath{\left[}}
\renewcommand{\]}{\ensuremath{\right]}}

\let\oldBracketLeft\{
\let\oldBracketRight\}
\renewcommand{\{}{\ensuremath{\left\oldBracketLeft}}
\renewcommand{\}}{\ensuremath{\right\oldBracketRight}}

\NewDocumentCommand\F{og}{%
  \IfNoValueTF{#1}%
    { \IfNoValueTF{#2}{\mathcal{F}}{\mathcal{F}\!\{{#2}\}} }%
    { \IfNoValueTF{#2}{\mathcal{F}\!\[{#1}\]}{\mathcal{F}\!\({#2}\)} }%
}
\NewDocumentCommand\FInv{og}{%
  \IfNoValueTF{#1}%
    { \IfNoValueTF{#2}{\mathcal{F}^{-1}}{\mathcal{F}^{-1}\!\{{#2}\}} }%
    { \IfNoValueTF{#2}{\mathcal{F}^{-1}\!\[{#1}\]}{\mathcal{F}^{-1}\!\({#2}\)} }%
}

\NewDocumentCommand\rect{g}{%
  \IfNoValueTF{#1}
  {\ensuremath{\mathrm{rect}}}
  {\ensuremath{\mathrm{rect}\of{#1}}}%
}

\NewDocumentCommand\sinc{g}{%
  \IfNoValueTF{#1}
  {\ensuremath{\mathrm{sinc}}}
  {\ensuremath{\mathrm{sinc}\of{#1}}}%
}

\NewDocumentCommand\supp{g}{%
  \IfNoValueTF{#1}
  {\ensuremath{\mathrm{supp}}}
  {\ensuremath{\mathrm{supp}\of{#1}}}%
}

\let\oldMathcal\mathcal
\renewcommand{\mathcal}[1]{\ensuremath{\oldMathcal{#1}}}

\makeatletter
\def\foreach#1#2#3{%
  \@test@foreach{#1}{#2}#3,\@end@token
}

\def\@swallow#1{}

\def\@test@foreach#1#2{%
  \@ifnextchar\@end@token%
    {\@swallow}%
    {\@foreach{#1}{#2}}%
}

\def\@foreach#1#2#3,#4\@end@token{%
  #1{#2}{#3}%
  \@test@foreach{#1}{#2}#4\@end@token%
}
\makeatother

\newtheorem{theorem}{Theorem}[section]
\newtheorem{lemma}[theorem]{Lemma}

\newtheorem{remark}{Remark}

\newenvironment{proof}[1][Proof]{\begin{trivlist}
\item[\hskip \labelsep {\bfseries #1}]}{\end{trivlist}}
\newenvironment{definition}[1][Definition]{\begin{trivlist}
\item[\hskip \labelsep {\bfseries #1}]}{\end{trivlist}}

\def\R{\mathbb R}
\def\l|{\left|}
\def\r|{\right|}
\def\l({\left(}
\def\r){\right)}
\def\l[{\left[}
\def\r]{\right]}

\renewcommand{\E}[1]{\mathbb{E}\left[ #1 \right]}
\newcommand{\half}{{\frac{1}{2}}}

\begin{document} 

\title{Direct Estimation of Information Divergence Using Nearest Neighbor Ratios}

\author[1]{Morteza Noshad\thanks{A.A@university.edu}}
\author[2]{Kevin R. Moon\thanks{B.B@university.edu}}
\author[1]{Salimeh Yasaei Sekeh\thanks{C.C@university.edu}}
\author[1,*]{Alfred O. Hero III\thanks{D.D@university.edu}}
\affil[1]{University of Michigan,
Electrical Engineering and Computer Science,
Ann Arbor, Michigan, U.S.A}
\affil[2]{Yale University,
Genetics and Applied Math Departments,
New Haven, Connecticut, U.S.A}

\renewcommand\Authands{ and }

\maketitle
\renewcommand{\thefootnote}{\fnsymbol{footnote}} \footnotetext[1]{This research was partially supported by ARO grant W911NF-15-1-0479.} 

\thispagestyle{fancy} 

\begin{abstract}
We propose  a direct estimation method for R\'{e}nyi and f-divergence measures based on a new  graph theoretical interpretation. Suppose that we are given two sample sets $X$ and $Y$, respectively with $N$ and $M$ samples, where $\eta:=M/N$ is a constant value. Considering the $k$-nearest neighbor ($k$-NN) graph of $Y$ in the joint data set $(X,Y)$, we show that the average powered ratio of the number of $X$ points to the number of $Y$ points among all $k$-NN points is proportional to R\'{e}nyi divergence of $X$ and $Y$ densities. A similar method can also be used to estimate f-divergence measures.  We derive bias and variance rates, and show that for the class of $\gamma$-H\"{o}lder smooth functions, the estimator achieves the MSE rate of $O\of{N^{-2\gamma/(\gamma+d)}}$. Furthermore, by using a weighted ensemble estimation technique, for density functions with continuous and bounded derivatives of up to the order $d$, and some extra conditions at the support set boundary, we derive an ensemble estimator that achieves the parametric MSE rate of $O(1/N)$.
Our estimator requires no boundary correction, and remarkably, the boundary issues do not show up. 
Our approach is also more computationally tractable than other competing estimators, which makes them appealing in many practical applications.

\end{abstract}
\section{Introduction}

Shannon entropy, mutual information, and the Kullback-Leibler (KL) divergence are major information theoretic measures. Shannon entropy can measure diversity or uncertainty of samples, while KL-divergence is a measure of dissimilarity, and mutual information is a measure of dependency between two  probability distributions \cite{cover2012}. R\'{e}nyi  proposed a divergence measure which generalizes KL-divergence \cite{rrnyi1961}. F-divergence is another general family which is also well studied, and comprises many important divergence measures such as KL-divergence, total variation distance, and $\alpha$-divergence \cite{ali}. 
These measures have wide range of applications in information and coding theory, statistics and machine learning \cite{cover2012,moon2014,structure2016}.

A major class of estimators for these measures is called non\hyp parametric, for which minimal assumptions on the density functions are considered in contrast to parametric estimators. An approach used for this class is plug-in estimation, in which we find an estimate of a distribution function and then plug it in the measure function. $k$-Nearest Neighbor ($K$-NN) and Kernel Density Estimator (KDE) methods are examples of this approach. Another approach is direct estimation, in which we find a relationship between the measure function and a functional in Euclidean space. 
In a seminal work in 1959, Beardwood et al derived the asymptotic behavior of the weighted functional of minimal graphs such as $K$-NN and TSP of $N$ i.i.d random points \cite{beardwood1959shortest}. They showed that the sum of weighted edges of these graphs converges to the integral of a weighted density function, which can be interpreted as R\'{e}nyi entropy. Since then, this work has been of great interest in signal processing and machine learning communities. More recent studies of direct graph theoretical approaches include the estimation of R\'{e}nyi entropy using the minimal graphs \cite{hero2003}, in which the authors investigate the convergence rates, as well as the estimation of Henze-Penrose divergence using MST graphs \cite{Henze}. Yet the extension to R\'{e}nyi divergence and f-divergences has remained an open question. Moreover, among various estimators of information measures, developing accurate and computationally tractable approaches has been often a challenge. Therefore, for  practical and computational reasons, direct graphical algorithms have been under attention in the literature including this work.

In this work, we propose an estimation method for R\'{e}nyi and f-divergences based on a direct graph estimation method. We show that given two sample sets $X$ and $Y$ with respective densities of $f_1$ and $f_2$, and the $k$-nearest neighbor ($k$-NN) graph of $Y$ in the joint data set $(X,Y)$, the average powered ratio of the number of $X$ points to the number of $Y$ points among all $k$-NN points converges to the R\'{e}nyi divergence. Using this fact, we design a consistent estimator for the R\'{e}nyi and f-divergences.

Unlike most distance-based divergence estimators, our proposed estimator can use non-Euclidean metrics,
which makes this estimator appealing in many information theoretic and machine learning applications. 
Our estimator requires no boundary correction, and surprisingly, the boundary issues do not show up. This is because the proposed estimator automatically cancels the extra bias of the boundary points in the ratio of nearest neighbor points. Our approach is more computationally tractable than other estimators, with a time complexity of $O(kN\log N)$, required to construct the $k$-NN graph \cite{vaidya1989}. For example for $k=N^{1/{d+1}}$ we get the complexity of $O(N^{(d+2)/(d+1)}\log N)$.
We show that for the class of $\gamma$-H\"{o}lder smooth functions, the estimator achieves the MSE rate of $O(N^{-2\gamma/(\gamma+d)})$.  Furthermore, by using the theory of optimally weighted ensemble estimation \cite{Kevin16,structure2016}, for density functions with continuous and bounded derivatives of up to the order $d$, and some extra conditions at the support set boundary, we derive an ensemble estimator that achieves the optimal MSE rate of $O(1/N)$, which is independent of the dimension.
Finally, the current work is an important step towards extending the direct estimation method studied in \cite{steele,yukich} to more general information theoretic measures.

Several previous works have investigated an estimator for a particular type of divergence measures. $k$-NN \cite{poczos2011}, KDE \cite{Poczos2014_2}, and histogram \cite{wang2009} estimators  are among the studied plug-in estimators for the f-divergence family. In general, most of these estimators suffer from several restrictions such as lack of analytic convergence rates, or high computational complexity. 

Recent works have focused on the MSE convergence rates for plug-in divergence estimators, such as KDE. Singh and P\'{o}czos proposed estimators for general density functionals and  R\'{e}nyi divergence, based on the kernel density plug-in estimator \cite{Poczos2014_2}\cite{Poczos2014_1}, which can achieve the convergence rate of $O(1/N)$ when the densities are at least $d$ times differentiable. 
In a similar approach, Kandasamy et al  proposed another KDE-based estimator for general density functionals and divergence measures, which can achieve the convergence rate of $O(1/N)$ when the densities are at least $d/2$ differentiable \cite{kandasamy}.

Moon et al proposed simple kernel density plug-in estimators using weighted ensemble methods to improve the rate \cite{Kevin16}\cite{kevin2014}. The proposed estimator can achieve the convergence rate when the densities are at least
$(d+1)/2$ times differentiable. 
The main drawback of these estimators is handling the bias at the support set boundary. For example, using the estimators proposed in \cite{Poczos2014_2,kandasamy} requires knowledge of the densities' support set and numerous computations at the support boundary, which become complicated when the dimension increases. To circumvent this issue, Moon et al \cite{Kevin16} assumed smoothness conditions at the support set boundary, which may not always be true in practice. In contrast, our basic estimator does not require any smoothness assumptions on the support set boundary although our ensemble estimator does. Regarding the algorithm time complexities, our estimator spends $O(kN\log N)$ time versus the time complexity of KDE based estimators which spend $O(N^2)$ time. 

A rather different method for estimating f-divergences is suggested by Nguyen et al \cite{Nguyen}, which is based on a variational representation of f-divergences that connects the estimation problem to a convex risk minimization problem. This approach achieves the parametric rate of $O(1/N)$ when the likelihood ratio is at least $d/
2$ times differentiable. However, the algorithm's time complexity is even worse than $O(N^2)$.


\section{A direct estimator of divergence measures}
In this section, we first introduce the R\'{e}nyi and f-divergence measures. Then we propose an estimator based on a graph theoretical interpretation, and we outline our main theoretical results, which will be proven in section \ref{Proof_Section}.

Consider two density functions $f_1$ and $f_2$ with  support $\mathcal{M}\subseteq \mathbb{R}^d$. The R\'{e}nyi divergence between $f_1$ and $f_2$ is
\begin{align}\label{def_Renyi_div}
D_{\alpha}\left(f_1(x)\vert\vert f_2(x)\right) & :=\frac{1}{\alpha-1}\log\int f_1(x)^\alpha f_2(x)^{1-\alpha}dx \nonumber\\
& = \frac{1}{\alpha-1}\log J_{\alpha}(f_1,f_2),
\end{align}
where in the second line, $J_{\alpha}(f_1,f_2)$ is defined as $J_{\alpha}(f_1,f_2):=\mathbb{E}_{f_2}\l[(\frac{f_1(x)}{f_2(x)})^\alpha\r]$:

Another general divergence family, f-divergence, is also defined as follows \cite{ali}.

\begin{align}\label{def_f_div}
D_g\left(f_1(x)\vert\vert f_2(x)\right) & :=\int g\of{\frac{f_1(x)}{f_2(x)}}f_2(x)dx \nonumber\\
& =\mathbb{E}_{f_2}\of[g\of{\frac{f_1(x)}{f_2(x)}}],
\end{align}
where $g$ is a smooth and convex function such that $g(1)=0$. KL-divergence, Hellinger distance and total variation distance are particular cases of this family. Note that for our approach, we only assume that $g$ is smooth.

We assume that the densities are lower bounded by $C_L>0$ and upper bounded by $C_U$. Also $f_1$ and $f_2$ belong to H\"{o}lder smoothness class with parameter $\gamma$:


\begin{definition}\label{Holder}
Given a support $\mathcal{X} \subseteq \mathbb{R}^d$, a function $f:\mathcal{X} \to \mathbb{R}$ is called H\"{o}lder continuous with parameter $0<\gamma\leq 1$, if there exists a positive constant $G_f$, depending on $f$, such that 
\begin{equation}
|f(y)-f(x)|\leq G_f\|y-x\|^{\gamma},
\end{equation}
for every $x\neq y \in \mathcal{X}$.
\end{definition}

The function $g(x)$ in \eqref{def_f_div} is also assumed to be Lipschitz continuous; i.e. $g$ is H\"{o}lder continuous with $\gamma=1$.


\begin{remark}
$\gamma$-H\"{o}lder smoothness family comprises a large class of continuous functions including continuously differentiable functions and Lipschitz continuous functions.
Also note that for $\gamma > 1$, any $\gamma$–H\"{o}lder continuous function on any bounded and continuous support is constant.  
\end{remark}

\begin{definition}
[Nearest Neighbor Ratio (NNR) Estimator: ] Consider the i.i.d samples $X=\{X_1,...,X_N\}$ drawn from $f_1$ and $Y=\{Y_1,...,Y_M\}$ drawn from $f_2$. We define the set $Z:=X\cup Y$, and consider the $k$-NN points for each of the points $Y_i$ in the set $Y$, which is represented by $Q_k(Y_i)$. Let $N_i$ and $M_i$ be the number of points of the sets $X$ and $Y$ among the $k$NN points of $Y_i$, respectively. Then an estimator for R\'{e}nyi divergence is

\begin{equation}\label{est_Renyi_def0}
\widetilde{D}_{\alpha}(X,Y):=\frac{1}{(\alpha-1)}\log \of[\frac{\eta^{\alpha}}{M}\sum_{i=1}^M\left(\frac{N_i}{M_i+1}\right)^\alpha],
\end{equation}
where $\eta:=M/N$. Similarly, using the alternative form in \eqref{def_Renyi_div}, we have 
\begin{equation}\label{est_J_def}
\widehat{J}_{\alpha}(X,Y):=\frac{\eta^{\alpha}}{M}\sum_{i=1}^M\left(\frac{N_i}{M_i+1}\right)^\alpha.
\end{equation}

Note that the estimator defined in \eqref{est_Renyi_def0} can be negative and unstable in extreme cases. To correct this, we propose the NNR estimator for R\'{e}nyi divergence denoted by $\widehat{D}_{\alpha}(X,Y)$:
\begin{equation}\label{est_Renyi_def}
\min\{\max\{\widetilde{D}_{\alpha}(X,Y),0\},\frac{1}{|1-\alpha|}\log\of{\frac{C_U}{C_L}}\}.
\end{equation}

The NNR f-divergence estimator is defined as
\begin{equation}\label{est_f_def}
\widehat{D}_{g}(X,Y):=\max\{\frac{1}{M}\sum_{i=1}^M \widetilde{g}\of{\frac{\eta N_i}{M_i+1}},0\},
\end{equation}
where $\widetilde{g}(x):=\max\{g(x),g\of{C_L/C_U}\}$.
\end{definition}

The intuition behind the proposed estimators is that, the ratio $\frac{N_i}{M_i+1}$ can be considered an estimate of density ratios at $Y_i$.  Note that if the densities $f_1$ and $f_2$ are almost equal, then for each point $Y_i$, $N_i \approx M_i+1$, and therefore both $\widehat{D}_{\alpha}(X,Y)$ and $\widehat{D}_{g}(X,Y)$ tend to zero. 
In the following theorems we derive upper bounds on the bias and variance rates. Consider the bias and variance definitions as $\mathbb{B}[\hat{T}]=\mathbb{E}[\hat{T}]-T$ and $\mathbb{V}[\hat{T}]=\mathbb{E}[\hat{T}^2]-\mathbb{E}[\hat{T}]^2$, respectively, where $\hat{T}$ is an estimator of the parameter $T$.

\begin{theorem} \label{bias_theorem}

 The bias of NNR estimator for R\'{e}nyi divergence,  defined in \eqref{est_Renyi_def}, can be bounded as
\begin{align} \label{bias_Renyi}
\mathbb{B}\of[\widehat{D}_{\alpha}(X,Y)]= O\of{\of{\frac{k}{N}}^{\gamma/d}}+ O\of{\frac{1}{k}}.
\end{align}
\end{theorem}
Here $\gamma$ is the H\"{o}lder smoothness parameter. 

\begin{theorem}\label{variance}
The variance of the NNR estimator is
\begin{align}
\mathbb{V}\of[\widehat{D}_{\alpha}(X,Y)]\leq O\of{\frac{1}{N}}+ O\of{\frac{1}{M}}.
\end{align}
\end{theorem}

\begin{remark}\label{variance_others}
The same variance bound holds true for the RV  $\widehat{J}_{\alpha}(X,Y)$. Also bias and variance results easily extend to the f-divergence estimator.
\end{remark}

\begin{remark}
Note that in most cases, the $1/k$ term in \eqref{bias_Renyi} is the dominant error term, and in order to have an asymptotically unbiased NNR estimator, $k$ should be a growing function of $N$. The $1/k$ term actually comes from the error of Poissonization technique used in the proof.
By equating the terms $O\of{k/N)^{\gamma/d}}$ and $O(1/k)$, it turns out that for $k_{opt}=O\of{N^{\frac{\gamma}{d+\gamma}}}$, we get the optimal MSE rate of $O\of{N^{\frac{-2\gamma}{d+\gamma}}}$. The optimal choice for $k$ can be compared to the optimum value $k=O\of{\sqrt{N}}$ in \cite{moon2014}, where a plug-in KNN estimator is used. Also considering the computational complexity of $O(kN\log N)$ to construct the $k$-NN graph \cite{vaidya1989}, we see that there is a trade-off between MSE rate and complexity for different values of $k$. In the particular case of optimal MSE, the computational complexity of this method is $O\of{N^{\frac{d+2\gamma}{d+\gamma}}\log N}$.
\end{remark}
\begin{algorithm} \label{algo}
\DontPrintSemicolon
\SetKwInOut{Input}{Input}\SetKwInOut{Output}{Output}
\Input{Data sets $X=\{X_1,...,X_N\}$, $Y=\{Y_1,...,Y_M\}$}

\BlankLine
 
$Z\leftarrow X \cup Y$    \;
\For {each point $Y_i$ in $Y$}{
		\tcc{Set of $k$-NN points of $Y_i$ in $Z$}
        $S_i\leftarrow \{Q_1(Y_i), ...,Q_k(Y_i) \}\qquad$ \\
		 $R_i \leftarrow |S_i\cap X|/|S_i \cap Y|$}
$\widehat{D} \leftarrow 1/(\alpha-1)\log\of[ \of{\eta^{\alpha}\sum_i R_i^\alpha}/M]$

\Output{$\widehat{D}$}

\caption{NNR Estimator of R\'{e}nyi Divergence }
\end{algorithm}

Under extra conditions on the densities and support set boundary, we can improve the bias rate by applying the ensemble theory in \cite{Kevin16,structure2016}. Assume that the density functions are in the H\"{o}lder space $\Sigma(\gamma,L)$, which consists of functions on $\mathcal{X}$ continuous derivatives up to order $q=\floor{\gamma}\geq d$  and the $q$th partial derivatives are H\"{o}lder continuous with exponent $\gamma'=:\gamma-q$. We also assume that the density derivatives up to order $d$ vanish at the boundary.
Let $\mathcal{L}:=\{l_1,...,l_L\}$ be a set of index values with $l_i<c$. Let $k(l):=\floor{l\sqrt{N}}$. The weighted ensemble estimator is defined as $\widehat{D}_w:=\sum_{l\in \mathcal{L}}w(l)\widehat{D}_{k(l)}$, where $\widehat{D}_{k(l)}$ is the NNR estimator of R\'{e}nyi $\alpha$-divergence, using the $k(l)$-NN graph. 

\begin{theorem} \label{ensemble_theorem}
Let $L>d$ and  $w_0$ be the solution to:
\begin{align}
\min_w &\qquad \|w\|_2 \nonumber\\
\textit{subject to} &\qquad \sum_{l\in \mathcal{L}}w(l)=1, \nonumber\\
&\qquad \sum_{l\in \mathcal{L}}w(l)l^{i/d}=0, i\in \mathbb{N}, i\leq d.
\end{align}
Then the MSE rate of the ensemble estimator $\widehat{D}_{w_0}$ is $O(1/N)$.
\end{theorem}

\section{Proof}\label{Proof_Section}

In this section we derive the bias terms of NNR estimator. The variance bound for NNR estimator is more straightforward and can be derived using Efron\hyp Stein inequality. Also for proving the MSE rate of ensemble variant of the NNR estimator, we need more accurate bias rates, which is provided in the arXiv version. So, for variance and ensemble estimation proofs we refer the reader to the Appendix section of arXiv version of the paper.
First, we provide a smoothness lemma for the densities. Unless stated otherwise, all proofs of lemmas are provided in the arXiv version.

\begin{lemma}\label{Holder_KNN}
Suppose that the density function $f(x)$ belongs to the $\gamma$-H\"{o}lder smoothness class. Then if $B(x,r)$ denotes the sphere with center $x$ and radius $r=\rho_k(x)$, where $\rho_k(x)$ is defined as the $k$-NN distance on the point $x$, we have the following smoothness condition:
\begin{align}\label{f_difference}
\mathbb{E}_{\rho_k(x)}\of[\sup_{y\in B(x,\rho_k(x))}\lvert f(y)-f(x)\rvert] \leq \epsilon_{\gamma,k},
\end{align}

where $O\of{(k/N)^{\gamma/d}}+O\of{\mathcal{C}(k)}$, and we have $\mathcal{C}(k):=exp(-3k^{1-\delta})$ for a fixed $\delta\in (2/3,1)$.
\end{lemma}

We first state the bias proof for R\'{e}nyi divergence, and then we extend the method to f-divergence. 
It is easier to work with $\widehat{J}_\alpha(X,Y)$ defined in \eqref{est_J_def}, instead of $\widehat{D}_\alpha(X,Y)$. The following lemma provides the essential tool to make a relation between $\mathbb{B}\of{\widehat{D}}$ and $\mathbb{B}\of{\widehat{J}}$.

\begin{lemma}\label{f2Z_bias}
Assume that $g(x): \mathcal{X}\to \mathbb{R}$ is Lipschitz continuous with constant $H_g>0$. If $\widehat{T}$ is a RV estimating a constant value $T$ with the bias $\mathbb{B}[\widehat{T}]$ and the variance $\mathbb{V}[\widehat{T}]$, then the bias of $g(\widehat{T})$ can be upper bounded by 
\begin{align}
\abs{\E{g(\widehat{T})-g(T)}} \leq H_g \of{\sqrt{\mathbb{V}\of[\widehat{T}]}+\abs{\mathbb{B}\of[\widehat{T}]}}.
\end{align}
\end{lemma}
An immediate consequence of this lemma is 
\begin{align} \label{DtoJ_bias}
&\abs{\mathbb{B}\of[\widehat{D}_{\alpha}(X,Y)]}\leq C\abs{\mathbb{B}\of[\widehat{J}_{\alpha}(X,Y)] + \sqrt{\mathbb{V}\of[\widehat{J}_{\alpha}(X,Y)]}},
\end{align}
where $C$ is a constant.

From theorem \ref{variance}, $\mathbb{V}\of[\widehat{J}_{\alpha}(X,Y)]=O(1/N)$, so we only need to bound $\mathbb{B}\of[\widehat{J}_{\alpha}(X,Y)]$.
If $\eta:=M/N$, we have:
\begin{align}\label{J2sum}
\E{\widehat{J}_{\alpha}(X,Y)} &=\frac{\eta^{\alpha}}{M}\E{\sum_{i=1}^M\left(\frac{N_i}{M_i+1}\right)^\alpha}\nonumber\\
&=\eta^{\alpha}\mathbb{E}_{Y_1\sim f_2(x)}\E{\left(\frac{N_1}{M_1+1}\right)^\alpha\middle\vert Y_1}.
\end{align}

Now note that $N_1$ and $M_1$ are not independent since $N_1+M_1=k$. We use the Poissonizing technique \cite{Barbour}\cite{depoisson} and assume that $N_1+M_1=K$, where $K$ is a Poisson random variable with mean $k$. We represent the Poissonized variant of $\widehat{J}_\alpha(X,Y)$ by $\overline{J}_\alpha(X,Y)$, and we will show that $\E{\widehat{J}_\alpha(X,Y)}=\E{\overline{J}_\alpha(X,Y)}+O(1/k)$.
By partitioning theorem for a Poisson random variable with Bernoulli trials of probabilities $\Pr\of{Q_i(Y_1)\in X}$ and $\Pr\of{Q_i(Y_1)\in Y}$, we argue that $N_1$ and $M_1$ are two independent Poisson RVs. 
We first compute $Pr\of{Q_k(Y_1)\in X}$ and $Pr\of{Q_k(Y_1)\in Y}$ as follows:

\begin{lemma}\label{point_probability}
Let $\eta:=M/N$. The probability that the point  $Q_k(Y_1)$ respectively belongs to the sets $X$ and $Y$ is equal to
\begin{align}\label{point_probability_equation}
\Pr\of{Q_k(Y_1)\in X} &= \frac{f_1(Y_1)}{f_1(Y_1)+\eta f_2(Y_1)} +O(\epsilon_{\gamma,k}) \nonumber\\
\Pr\of{Q_k(Y_1)\in Y} &= \frac{\eta f_2(Y_1)}{f_1(Y_1)+\eta f_2(Y_1)} +O(\epsilon_{\gamma,k}).
\end{align}
\end{lemma}

Using the conditional independence of $N_1$ and  $M_1$ we write 
\begin{align}\label{E_ratio}
&\E{\frac{N_1}{M_1+1}\middle\vert Y_1} = \E{N_1\middle\vert Y_1} \E{(M_1+1)^{-1}\middle\vert Y_1}.
\end{align}
 $\E{N_1\vert Y_1}$ can be simplified as 
\begin{align}
\E{N_1\vert Y_1}&=\sum_{i=1}^k\Pr\of{Q_i(Y_1)\in X}\nonumber\\
&=k\frac{f_1(Y_1)}{f_1(Y_1)+\eta f_2(Y_1)} +O(k\epsilon_{\gamma,k}).
\end{align}
Also similarly, $$\E{M_1\vert Y_1}=\frac{k\eta f_2(Y_1)}{f_1(Y_1)+\eta f_2(Y_1)} +O(k\epsilon_{\gamma,k}).$$

\begin{lemma}\label{lemma_poisson}
If $U$ is a Poisson random variable with the mean $\lambda>1$, then 
\begin{equation}
\E{(U+1)^{-1}} = \frac{1}{\lambda}\of{1-e^{-\lambda}}.
\end{equation}

\end{lemma}
Using this lemma for $M_1$ yields
\begin{align}\label{E_M}
&\E{(M_1+1)^{-1} \middle\vert Y_1}  \nonumber\\ 
&\qquad=k^{-1}\l[\frac{\eta f_2(Y_1)}{f_1(Y_1)+\eta f_2(Y_1)} +O(\epsilon_{\gamma,k})\r]^{-1}+O\of{\frac{e^{-vk}}{k}},
\end{align}
here $v$ is some positive constant.
Therefore, \eqref{E_ratio} becomes
\begin{align}\label{FLRWT_proof_5}
\E{\frac{N_1}{M_1+1}\middle\vert Y_1} = \frac{f_1(Y_1)}{\eta f_2(Y_1)}+O(\epsilon_{\gamma,k})+O\of{e^{-vk}}.
\end{align}
Using lemma \ref{f2Z_bias} and theorem \ref{variance}, we obtain
 \begin{align}\label{Expec_ratio}
&\E{\left(\frac{N_1}{M_1+1}\right)^\alpha\middle\vert Y_1} = \eta^{-\alpha}\left(\frac{f_1(Y_1)}{f_2(Y_1)}\right)^{\alpha}+\qquad\nonumber\\
& \qquad\qquad +O(\epsilon_{\gamma,k})+O\of{e^{-vk}}+O(N^{-\half}).
\end{align}
By applying an equation similar to \eqref{J2sum}, we get
\begin{align} \label{bias_J_bar}
\mathbb{B}\of[\overline{J}_\alpha(X,Y)]=O(\epsilon_{\gamma,k})+O\of{e^{-vk}}+O(N^{-\half}).
\end{align}

\begin{lemma}\label{depoissonize}
De-Poissonizing $\overline{J}_\alpha(X,Y)$ adds $O(\frac{1}{k})$ error:
\begin{align}
\E{\widehat{J}_\alpha(X,Y)}=\E{\overline{J}_\alpha(X,Y)}+O(1/k).
\end{align}
\end{lemma}

At this point the bias proof of NNR estimator for R\'{e}nyi divergence is complete, and since $O\of{e^{-vk}}$ and $O\of{N^{-\half}}$ are of higher order compared to $O\of{\epsilon_{\gamma,k}}$, we obtain the final bias rate in \eqref{bias_Renyi}. The bias proof of NNR estimator for f-divergence is similar, and by using the lemma \ref{f2Z_bias} for $g$, we can follow the same steps to prove the bias bound. The complete proof is provided in the arXiv version.

\section{numerical Results}

In this section we provide  numerical results to show the consistency of the proposed estimator and compare the estimation quality in terms of different parameters such as $N$ and $k$. 
In our experiments, we choose i.i.d samples for $X$ and $Y$ from different independent distributions such as Gaussian, truncated Gaussian and uniform functions.

\begin{figure}
	\centering
	\includegraphics[width=9cm]{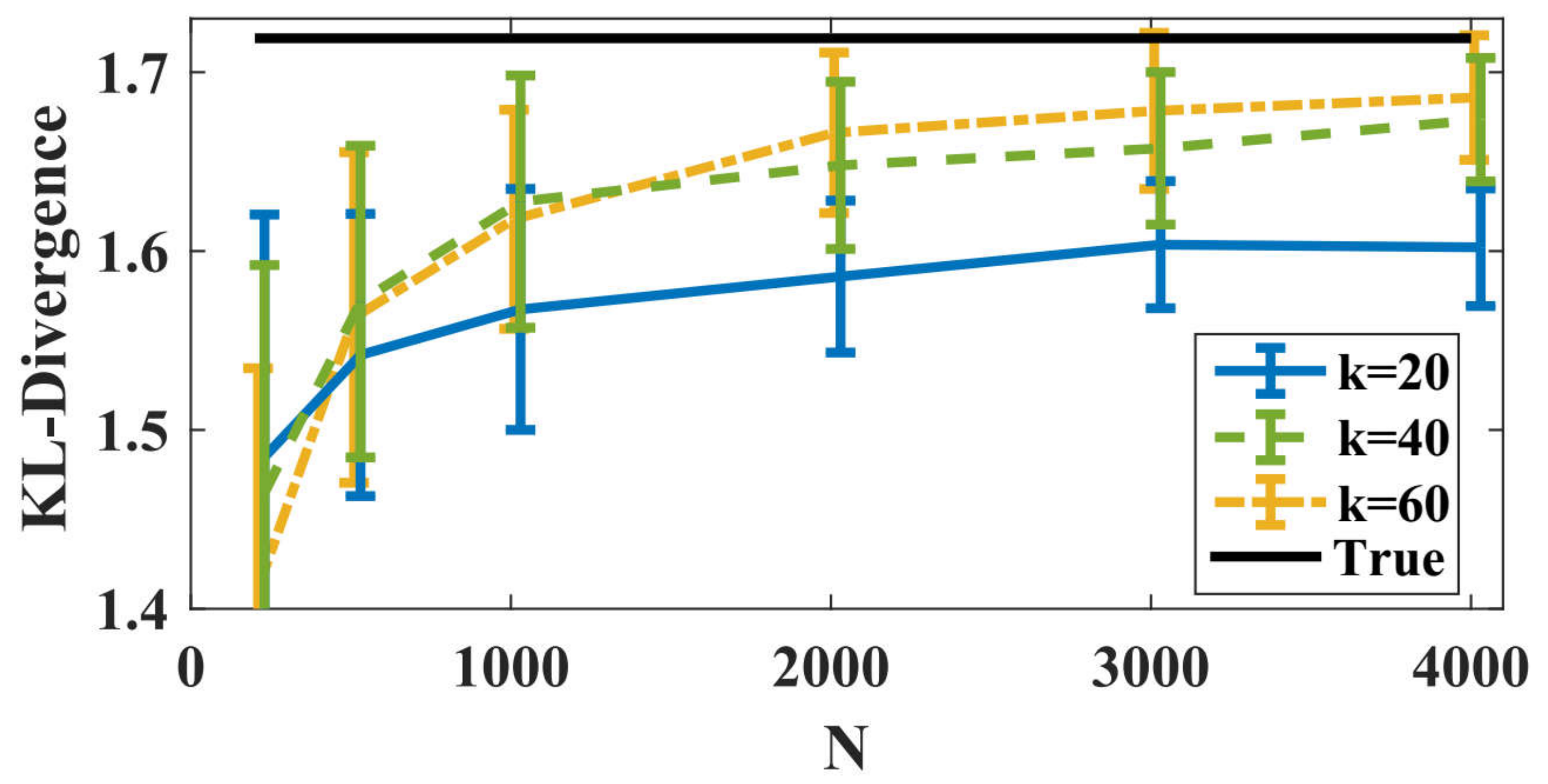}
	\caption{The estimated value for various values of $k$ is compared with the true value for KL-divergence between a Gaussian and a uniform distribution}\label{figure1}
\end{figure}

The first experiment, shown in Figure \ref{figure1}, shows the mean estimated KL-divergence as N grows for $k$ equal to $20,40,60$. The divergence measure is between a 2D Gaussian RV with mean $[0, 0]$ and variance of $2I_2$, and a uniform distribution with $x,y\in [-1,1]$.  For each case we repeat the experiment $100$ times, and compute the mean  of the estimated value and the standard deviation error bars. For small sample sizes, smaller $k$ results in smaller bias error, which is due to the $\of{\frac{k}{N}}^{\gamma/d}$ bias term. As $N$ grows, we get larger bias for small values of $k$, which is due to the fact that the $\of{1/k}$ term dominates. If we compare the standard deviations for different values of $k$ at $N=4000$, they are almost equal, which verifies the fact that variance is independent of $k$.

\begin{figure}
	\centering
	\includegraphics[width=9.0cm]{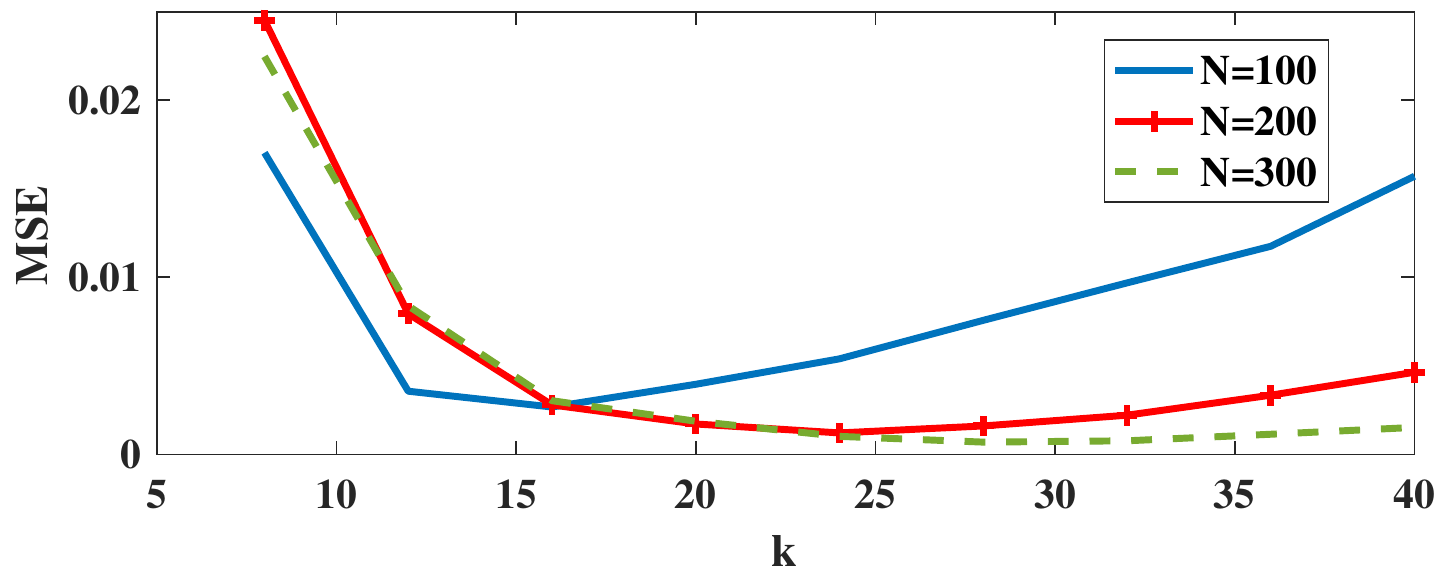}
	\caption{MSE of NNR estimator of R\'{e}nyi divergence with $\alpha=0.5$ for two independent, truncated normal RVs, as a function of $k$.}\label{optimumk}
\end{figure}

Figure \ref{optimumk} shows the MSE of NNR estimator of Renyi divergence with $\alpha=0.5$ for two independent, truncated normal RVs. The RVs are 2D with means $\mu_1=\mu_2=[0, 0]$ and covariance matrices $\sigma_1=I_2$ and $\sigma_2=3I_2$, where $I_2$ is a diagonal matrix of size $2$. Both of the RVs are truncated with the range $x\in [-2, 2]$ and $y\in [-2, 2]$.
In this figure we show the MSE for three different sample sizes of $100,200$, and $300$ for different values of $k$.
As $k$ increases initially, MSE decreases due to the $O(1/k)$ bias term. After reaching an optimal point, MSE increases as $k$ increases, indicating that the other bias terms begin to dominate. The optimal $k$ increases with the sample size which validates our theory.

\begin{figure}
	\centering
	\includegraphics[width=9cm]{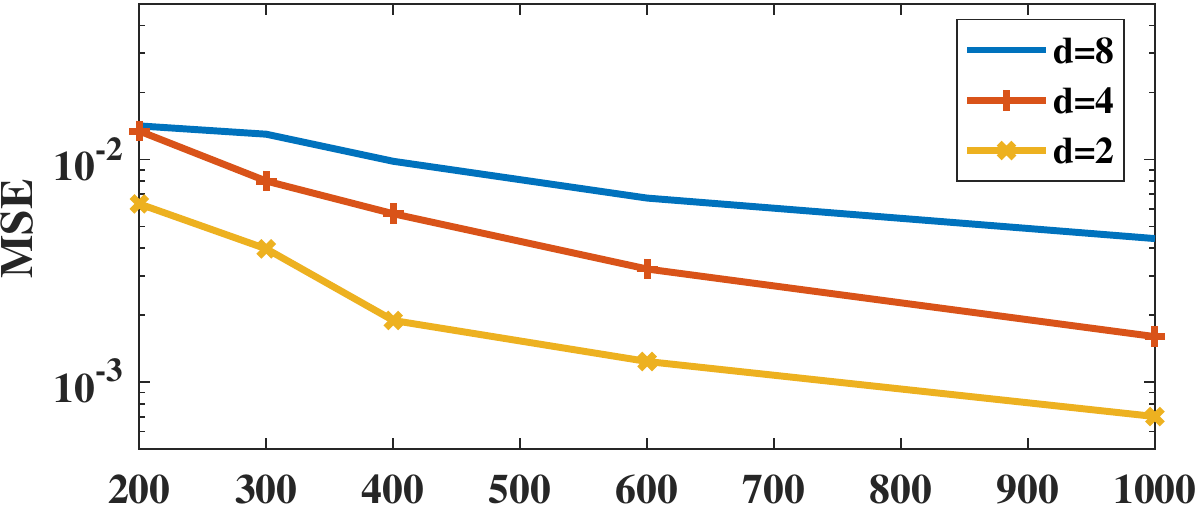}
	\caption{MSE of NNR estimator of Renyi divergence with $\alpha=2$ versus $N$, for two i.i.d. Normal RVs.}\label{d}
\end{figure}

Figure \ref{d} shows the MSE of the NNR estimator of R\'{e}nyi divergence with $\alpha=2$ versus $N$, for two i.i.d. Normal RVs for three different dimension sizes: $2,4$, and $8$. $k=90$ is fixed so that the $O\of{1/k}$ term in the bias can be ignored relative to the $O\of{(k/N)^{\gamma/d}}$ term. As dimension grows, the MSE decreases almost linearly in the logarithmic scale, which verifies the $O\of{(k/N)^{\gamma/d}}$ bias term.

\begin{figure}
	\centering
	\includegraphics[width=9cm]{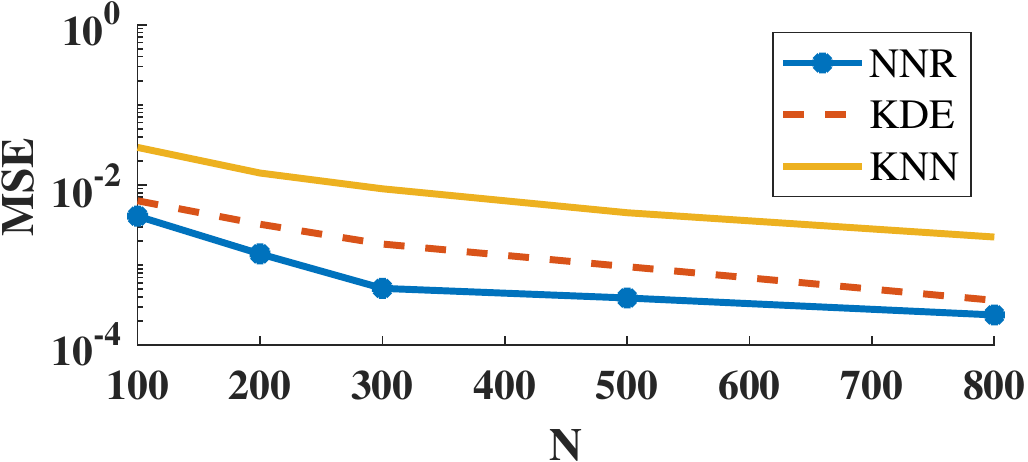}
    \caption{Comparison of MSE for estimation of R\'{e}nyi divergence ($\alpha=0.5$) between two Normal RVs the same mean $[0, 0]$ and different variances of $\sigma_1^2=I_2, \sigma_2^2=3I_2$ using NNR, KDE and $K$NN estimators.}\label{compare}
\end{figure}

Finally in Figure \ref{compare}, we compare our estimator with two standard plug-in estimators, $k$-NN, KDE. For each of these estimators we estimate the density at each $y\in Y$, and then compute the relation for the divergence measure using the definition in \eqref{def_Renyi_div}.  The graph shows the MSE  for R\'{e}nyi divergence ($\alpha=0.5$) between two Gaussian random variables with the same mean and different variances ($\sigma_1^2=I_2, \sigma_2^2=3I_2$) as a function of sample size, $N$. For both the NNR and $k$-NN estimators we use the optimal value for $k$ and the optimal bandwidth for the KDE estimator. According to this figure, the NNR estimator outperforms the other methods.

\section{Conclusion}
In this paper we proposed  a direct estimation method for R\'{e}nyi and f-divergence measures based on a new  graph theoretical interpretation. We proved bias and variance convergence rates, and validated our results by numerical experiments. Direct estimation procedures that converge for a fixed number $k$ of nearest neighbors is a worthwhile topic for future work.


\bibliographystyle{ieeetr}
\bibliography{citations.bib}
\onecolumn
\newpage
\section*{A. Bias Proof}
In this section we give proofs for the Lemmas \ref{Holder_KNN}, \ref{f2Z_bias}, \ref{point_probability}, \ref{lemma_poisson} and  \ref{depoissonize}. 

For proving Lemma \ref{Holder_KNN}, we need to derive a bound on the moments of $k$-NN distances. We define the $k$-NN ball centered at $x$ as
\begin{align}
S_k(x) := \{y : d(x, y) \leq \rho_k(x)\}.
\end{align}
Let $\textbf{V}_{k,N}(x)$ denote 
the volume of the $k$-NN ball with $N$ samples. Set

\begin{align}
\alpha_k(x):=\frac{\int_{S_k(x)\cap \mathcal{X}} dz}{\int_{S_k(x)} dz}.
\end{align}

Let $\mathcal{X_I}$ and $\mathcal{X_B}$ respectively denote the interior support and boundary of the support. For a point $x\in \mathcal{X}_I$ we have $\alpha_k=1$, and for $x\in \mathcal{X_B}$ we have $\alpha_k<1$. Note that the definition of interior and boundary points depends on $k$ and $N$.

\begin{lemma}\label{KNN_bias_Lemma}
We have the following relation for any $t \in \mathbb{R}$ and for each point $x\in \mathcal{X_I}$ with density $f(x)$:

\begin{align}\label{KNN_bias}
\E{\rho_k^t(x)}=\of{\frac{k}{c_dNf(x)}}^{t/d}+O\of{\frac{N^{-t/d}}{k}} +u(x)O\of{\of{\frac{k}{N}}^{t/d+2}}+o\of{\of{\frac{k}{N}}^{t/d+2}}+O\of{\of{\frac{k}{N}}^{t/d}\mathcal{C}(k)},
\end{align}

where $u(x)=g'(f(x))h(x)$ ,and $h$ is some bounded function of the density which is defined in \cite{sricharan2012}. 
\end{lemma}

\begin{proof}
We start with a result from \cite{sricharan2012}, A.25. Let $g:\R^+\to \R$ be some arbitrary function, then we have the following relation

\begin{align}
\E{g\left(\frac{k}{c_0n\rho_k^d(x)}\right)}=g(f(x))g_1(k,N)+g_2(k,N)+g'(f(x))h(x)(k/N)^2+o((k/N)^2)+O(\mathcal{C}(k)).
\end{align}
 where $g_1$ and $g_2$ are bias correction functions which depend on $g$. We also have $\mathcal{C}(k):=exp(-3k^{1-\delta})$ for a fixed $\delta\in (2/3,1)$. For example, if we set $k=(\log(N))^{1/(1-\delta)}$, then $O(\mathcal{C}(k))=O(1/N^3)$. Note that this term is negligible compared to other bias terms in our work.
 
Now  according to \cite{sricharan2012}, if we set $g(x)=x^{-\beta}$, then we have  $g_1(k,N)=\frac{\Gamma(k)}{\Gamma(k-\beta)(k-1)^{\beta}}$ and $g_2(k,N)=0$, which yields 

\begin{align}
\E{\rho_k^t(x)}=f(x)^{-t/d}\frac{\Gamma(k)}{\Gamma(k-t/d)}c'_0N^{-t/d} +u(x)O((\frac{k}{N})^{t/d+2})+o((\frac{k}{N})^{t/d+2})+O((\frac{k}{N})^{t/d}\mathcal{C}(k)).
\end{align}
Finally, using the approximation $\frac{\Gamma(k)}{\Gamma(k-\beta)}=k^\beta +O(1/k)$ results in \eqref{KNN_bias}.
\end{proof}


Now for the case of a bounded support, we derive an upper bound on $k$-NN distances for the points at the boundary:

\begin{lemma}\label{KNN_bias_boundary_Lemma}
For every point $x\in \mathcal{X_B}$ and any $t\in \mathbb{R}$ we have
\begin{align}
\E{\rho_k^t(x)}=O\of{(k/N)^{t/d}}+O\of{\mathcal{C}(k)}.
\end{align}
\end{lemma}

\begin{proof}

Define ${V}_{k,N}(x):= \frac{k}{N\alpha_k(x)f(x)}$. Let $p(k,N)$ denote any positive function satisfying $p(k,N)=\Theta\of{(k/N)^{2/d}}$ +$\frac{\sqrt{6}}{k^{\delta/2}}$ for some $\delta>0$. Further consider the event $E_1$ as

\begin{align}\label{Event}
E_1:=\{\abs{\frac{\textbf{V}_{k,N}(X)}{{V}_{k,N}(X)}-1} > p(k,N)\},
\end{align}
and $E_2$ as its complementary event. By using (B.2) in \cite{sricharan2012} (Appendix B), we have

\begin{align} \label{P_event}
Pr\of{E_1}=O\of{\mathcal{C}(k)}.
\end{align}

Moreover, we can simplify \eqref{Event} as:

\begin{align}\label{event_simplified}
\abs{c_d\rho_k^d(x)-\frac{k}{N\alpha_k(x)f(x)}} > \frac{kp(k,N)}{N\alpha_k(x)f(x)}.
\end{align}

Further we write $\E{\rho_k^\gamma(x)}$ as the sum of conditional expectations:

\begin{align} \label{bound_rho}
\E{\rho_k^\gamma(x)} &= \E{\rho_k^\gamma(x)\vert E_1} Pr\of{E_1}+\E{\rho_k^\gamma(x)\vert {E}_2} Pr\of{{E}_2} \nonumber\\
&=O\of{\mathcal{C}(k)}+\E{\rho_k^\gamma(x)\vert E_2} \of{1- O\of{\mathcal{C}(k)}} \nonumber\\
&=O\of{(k/N)^{\gamma/d}}+O\of{\mathcal{C}(k)},
\end{align}
where in the second line we have used \eqref{P_event} and also the fact that $\rho_k(x)$ is bounded from above because of the bounded support.

\end{proof}


\begin{proof}[\textbf{Proof of Lemma \ref{Holder_KNN}}]

From definition of Holder smoothness, for every $y\in B(x,\rho_k(x))$ we have

\begin{equation}
|f(y)-f(x)|\leq G_f\|y-x\|^{\gamma} \leq G_f\rho^\gamma_k(x).
\end{equation}

Using Lemmas \ref{KNN_bias_Lemma} and \ref{KNN_bias_boundary_Lemma} results in

\begin{align}
\mathbb{E}_{\rho_k(x)}\of[\sup_{y\in B(x,\rho_k(x))}\lvert f(y)-f(x)\rvert] \leq \epsilon_{\gamma,k},
\end{align}
where $O\of{(k/N)^{\gamma/d}}+O\of{\mathcal{C}(k)}$.
Note that all other terms in \eqref{KNN_bias} are of higher order and can be ignored.
\end{proof}


\begin{proof}[\textbf{Proof of lemma \ref{f2Z_bias}}]

\begin{align}
\abs{\E{g\of{\widehat{Z}}-g(Z)}} & \leq \abs{\E{g\of{\widehat{Z}}-g\of{\E{\widehat{Z}}}}} + \abs{\E{g\of{\E{\widehat{Z}}}-g(Z)}}  \nonumber\\
& \leq \E{\abs{g\of{\widehat{Z}}-g\of{\E{\widehat{Z}}}}} + H_g\abs{\E{\widehat{Z}}-Z} \nonumber\\
& \leq H_g \E{\abs{\widehat{Z}-\E{\widehat{Z}}}} + H_g\abs{\E{\widehat{Z}}-Z} \nonumber\\
& \leq H_g \of{\sqrt{\mathbb{V}\of[\widehat{Z}]}+\abs{\mathbb{B}\of[\widehat{Z}]}}.
\end{align}

In the second line we have used triangle inequality for the first term, and Lipschitz condition for the second term. Again in the third line, we have applied Lipschitz condition for the first term, and finally in the forth line we have used Cauchy\hyp Schwarz inequality.

\end{proof}

\begin{proof} [\textbf{Proof of Lemma \ref{point_probability}}]

Consider the following lemma which is proved immediately after the proof of Lemma \ref{point_probability} :

\begin{lemma}\label{general_P_QinX_lemma}
Let for any point $y\in \mathcal{X}$ define $\xi_1(y):=f_1(y)-f_1(Y_1)$ and $\xi_2(y):=f_2(y)-f_2(Y_1)$. Then $\Pr\of{Q_k(Y_1)\in X}$ can be derived as

\begin{align}\label{P_to_ratio}
\Pr\of{Q_k(Y_1)\in X}=\frac{f_1(Y_1)}{f_1(Y_1)+\eta f_2(Y_1)}+\tau_1(Y_1)+\tau_2(Y_1), 
\end{align}
where $\tau_1(Y_1)$ and $\tau_2(Y_1)$ are defined as

\begin{align}\label{P_to_ratio_errors}
\tau_1(Y_1)&:=\of{f_1(Y_1)+\eta f_2(Y_1)}^{-1}\mathbb{E}_{y\sim f_{Q_k(Y_1)}}\of[\xi_1(y)]\nonumber\\
\tau_2(Y_1) &:= \mathbb{E}_{y\sim f_{Q_k(Y_1)}}\of[\of{\frac{f_1(Y_1)}{f_1(Y_1)+\eta f_2(Y_1)}+\frac{\xi_1(y)}{f_1(Y_1)+\eta f_2(Y_1)}}\mathcal{U}\of{\frac{\xi_1(y)+\eta\xi_2(y)}{f_1(Y_1)+\eta f_2(Y_1)}}],
\end{align}
and $\mathcal{U}\of{x}:=1+\sum_{i=1}^{\infty}(-1)^i\of{x}^i$.
\end{lemma}

Now from Lemma \ref{Holder_KNN} we can simply write $\tau_1(Y_1)=O\of(\epsilon_{\gamma,k})$ and $\tau_2(Y_1)=O\of(\epsilon_{\gamma,k})$ which results in:

\begin{align}
\Pr\of{Q_k(Y_1)\in X} &= \frac{f_1(Y_1)}{f_1(Y_1)+\eta f_2(Y_1)} +O(\epsilon_{\gamma,k}).
\end{align}

\begin{remark}
It can similarly be proven that

\begin{align}
\Pr\of{Q_k(Y_1)\in Y} &= \frac{\eta f_2(Y_1)}{f_1(Y_1)+\eta f_2(Y_1)} +O(\epsilon_{\gamma,k}).
\end{align}

\end{remark}

\end{proof}

\begin{proof}[\textbf{Proof of Lemma \ref{general_P_QinX_lemma}}]
Let $B(Q_k(Y_1),\epsilon)$ be the sphere with the center $Q_k(Y_1)$ (the $k$-NN point of $Y_1$) and some small radius $\epsilon>0$. Also let $E_X$ and $E_Z$ denote the following events:

\begin{align}
E_X &:= \{\exists x\in X \mid x\in B(Q_k(Y_1),\epsilon)\}, \nonumber\\
E_Z &:= \{\exists x\in Z \mid x\in B(Q_k(Y_1),\epsilon)\}.
\end{align}

Let use the notation $\Pr\of{E_X(y)}$ to denote $\Pr\of{E_X|Q_k(Y_1)=y}$.

Suppose $f_{Q_k(Y_1)}$ be the density function of the RV $Q_k(Y_1)$. Then  $Pr\of{Q_k(Y_1)\in X}$ can be written as:

\begin{align}\label{P_QinX}
Pr\of{Q_k(Y_1)\in X}=\int_{\mathcal{X}}f_{Q_k(Y_1)}(y)Pr\of{Q_k(Y_1)\in X|Q_k(Y_1)=y},
\end{align}
where $Pr\of{Q_k(Y_1)\in X|Q_k(Y_1)=y}$ can be formulated using $E_X(y)$ and $E_Y(y)$ as

\begin{align}\label{P_E_ratio}
Pr\of{Q_k(Y_1)\in X|Q_k(Y_1)=y}=\frac{\Pr\of{E_X(y)}}{\Pr\of{E_Z(y)}}.
\end{align}

Let $P_f\of{y,\epsilon}$ denote the  probability of the sphere $B\of{y,\epsilon}$ with density $f$. Then there exist a function real function $\Delta_1(\epsilon)$ such  that for any $\epsilon>0$ we have

\begin{align}\label{Delta_1}
P_f\of{y,\epsilon}=f(y)c_d\epsilon^d+\Delta_1(\epsilon),
\end{align}

where $c_d$ is volume of the unit ball in dimension $d$. From definition of the density function we have 
\begin{align}\label{Delta_2}
f(y)=\lim_{\epsilon\to 0}\frac{P_f(y,\epsilon)}{c_d\epsilon^d}.
\end{align}

So, from \eqref{Delta_1} and \eqref{Delta_2} we get $\lim_{\epsilon\to 0}\Delta_1(\epsilon)/\epsilon^d=0$.

Now we compute $Pr(E_X(y))$ as

\begin{align}\label{P_EX}
\Pr(E_X(y))&=1-\of{1-P_{f_1}\of{y,\epsilon}}^N \nonumber\\
&=NP_{f_1}\of{y,\epsilon}+\Delta_1(\epsilon)+\sum_{i=2}^N (-1)^i{N\choose i}P_{f_1}\of{y,\epsilon}^i\nonumber\\
&=N c_d{f_1}\of{y}\epsilon^d+\Delta_2(\epsilon),
\end{align}
where $\Delta_2(\epsilon):=\Delta_1(\epsilon)+\sum_{i=2}^N (-1)^i{N\choose i}P_{f_1}\of{y,\epsilon}^i$. Note that $\lim_{\epsilon\to 0}\Delta_2(\epsilon)/\epsilon^d=0$.

Similarly, for $Pr(E_z)$ we can prove that

\begin{align}\label{P_EZ}
\Pr(E_z)=N c_d{f_1}\of{y}\epsilon^d+M c_d{f_2}\of{y}\epsilon^d+\Delta'_2(\epsilon),
\end{align}
where $\Delta'_2(\epsilon)$ is a function satisfying $\lim_{\epsilon\to 0}\Delta'_2(\epsilon)/\epsilon^d=0$.

From \eqref{P_E_ratio}, and considering the fact that \eqref{P_EX} and \eqref{P_EZ} hold true for any $\epsilon>0$, we get

\begin{align}\label{P_QinX_cond}
Pr\of{Q_k(Y_1)\in X|Q_k(Y_1)=y}=\lim_{\epsilon\to 0}\frac{\Pr\of{E_X(y)}}{\Pr\of{E_Z(y)}}= \frac{f_1(y)}{f_1(y)+\eta f_2(y)},
\end{align}

where $\eta=M/N$. Considering the Taylor expansion of $\frac{A+a}{B+b}$ for any real number $A,B,a,b$ such that $a\ll A$ and $b\ll B$, we have

\begin{align}\label{ratio_taylor}
\frac{A+a}{B+b}=\of{\frac{A}{B}+\frac{a}{B}}\of{1+\sum_{i=1}^{\infty}(-1)^i\of{\frac{b}{B}}^i}=\frac{A}{B}+\frac{a}{B}+\of{\frac{A}{B}+\frac{a}{B}}\mathcal{U}\of{\frac{b}{B}},
\end{align}
where $\mathcal{U}\of{x}:=\sum_{i=1}^{\infty}(-1)^i\of{x}^i$.
Consequently, by using this fact and relation \eqref{P_QinX_cond} we have

\begin{align}
Pr\of{Q_k(Y_1)\in X}&=\int_{\mathcal{X}}f_{Q_k(Y_1)}(y)\frac{f_1(y)}{f_1(y)+\eta f_2(y)} dy \nonumber\\
&=\frac{f_1(Y_1)}{f_1(Y_1)+\eta f_2(Y_1)}+\tau_1(Y_1)+\tau_2(Y_1), 
\end{align}
and $\tau_1(Y_1)$ and $\tau_2(Y_1)$ are given by

\begin{align}
\tau_1(Y_1)&=\of{f_1(Y_1)+\eta f_2(Y_1)}^{-1}\mathbb{E}_{y\sim f_{Q_k(Y_1)}}\of[\xi_1(y)]\nonumber\\
\tau_2(Y_1) &= \mathbb{E}_{y\sim f_{Q_k(Y_1)}}\of[\of{\frac{f_1(Y_1)}{f_1(Y_1)+\eta f_2(Y_1)}+\frac{\xi_1(y)}{f_1(Y_1)+\eta f_2(Y_1)}}\mathcal{U}\of{\frac{\xi_1(y)+\eta\xi_2(y)}{f_1(Y_1)+\eta f_2(Y_1)}}].
\end{align}

\end{proof}


\begin{proof}[\textbf{Proof of \ref{lemma_poisson}}]
From definition of Poisson RV, we can write

\begin{align}
\E{(U+1)^{-1}} = \sum_{k=0}^\infty \frac{1}{k+1}\of{\frac{\lambda^ke^{-\lambda}}{k!}} = \frac{1}{\lambda}\sum_{k=0}^\infty \frac{\lambda^{k+1}e^{-\lambda}}{(k+1)!} = \frac{1}{\lambda}\of{1-e^{-\lambda}}.
\end{align}

\end{proof}


\begin{proof}[\textbf{Proof of \ref{depoissonize}}]

We use the following theorem from \cite{depoisson} to de-possonize the estimator.

\begin{theorem}\label{depoissonize_thm}
Assume a sequence $a_n$ is given, and its poisson transform is $F(Z)$:

\begin{equation}
F(z)=\sum_{n\geq 0} a_n \frac{z^n}{n!}e^{-z}.
\end{equation}

Consider a linear cone $S_{\theta}=\{z:\abs{\arg(z)}\leq \theta, \theta <\pi/2\}$. Let the following conditions hold for some constants $R>0$, $\alpha <1$ and $\beta\in \mathbb{R}$:

\begin{itemize}
\item For $z\in S_\theta$, 
\begin{equation}
\abs{z}>R \Rightarrow |F(z)| = O\of{z^\beta}.
\end{equation}

\item For $z\notin S_\theta$, 
\begin{equation}
\abs{z}>R \Rightarrow |F(z)e^z| = O\of{e^{\alpha|z|}}.
\end{equation}

\end{itemize}

Then we have the following expansion that holds for every fixed $m$:

\begin{align} \label{depoisson_F2a}
a_n=\sum_{i=0}^m\sum_{j=0}^{i+m} b_{ij}n^i F^{(j)}(n)+O(n^{\beta-m-1/2}),
\end{align}
where $\sum_{ij}b_{ij}x^iy^j=\exp\of{x\log(1+y)-xy}$.

Let $\widehat{J}_{\alpha,k}(X,Y)$ and $\overline{J}_{\alpha,k}(X,Y)$ respectively represent the RVs $\widehat{J}_{\alpha}(X,Y)$ and $\overline{J}_{\alpha}(X,Y)$ with the parameter $k$.

Using the dePoissonization theorem, we take $a_k:=\E{\widehat{J}_{\alpha,k}(X,Y)}$ and $F(k):=\E{\overline{J}_{\alpha,k}(X,Y)}$. 
Since we are only interested in the values of $k$, for which $\lim_{N\to \infty}\frac{k}{N}=0$, we can assume $F(z)=O(1)$. So, both the first and second conditions of the Theorem \ref{depoissonize_thm} are satisfied. Then from \eqref{depoisson_F2a}, for $m=1$:

\begin{align}
\E{\widehat{J}_{\alpha,k}(X,Y)}= \E{\overline{J}_{\alpha,k}(X,Y)}+O\of{\frac{1}{k}}+\half O\of{\frac{1}{k^2}}+O\of{k^{-{3/2}}},
\end{align}
where $\beta=0$.
\end{theorem}

\end{proof}

Finally at the end of this section, we mention that the bias proof for $\widehat{D}_g(X,Y)$ is pretty similar to the bias proof of $\widehat{D}_g(X,Y)$ and simply follows by the same steps.


\section*{B. Ensemble Estimator}
In this section we state the MSE proof of the ensemble estimator.
Assume that the density functions are from the H\"{o}lder space $\Sigma(\gamma,L)$, which consists of those functions on $\mathcal{X}$ having continuous derivatives up to order $q$ and the $q$th partial derivatives are H\"{o}lder continuous with exponent $\gamma'$, where $q:=\floor{\gamma}$ and $\gamma':=\gamma-q$. We first compute the bias of interior points, by providing the following lemma.

\begin{lemma}\label{epsilon_ensemble}
For a constant parameter $\kappa\in \mathbb{N}$, let define $\mathcal{X_I^\kappa}:=\{x|x\in\mathcal{X}, \alpha_{\kappa}(x)=1 \}$ and $\mathcal{X_B^\kappa}:=\{x|x\in\mathcal{X}, \alpha_{\kappa}(x)<1 \}$. Then for any point $Y_1\in \mathcal{X}$ and any $k\leq \kappa$ we have
\begin{align}
&\E{\left(\frac{N_1}{M_1+1}\right)^\alpha\middle\vert Y_1} = \eta^{-\alpha}\left(\frac{f_1(Y_1)}{f_2(Y_1)}\right)^{\alpha} +\theta_\gamma(Y_1)+O\of{e^{-vk}}+O(N^{-\half}),
\end{align}
where $v$ is a constant defined in Lemma \ref{lemma_poisson} and $\theta_\gamma(Y_1)$ is given by

\begin{align}
\theta_\gamma(Y_1):=
\begin{cases} 
      \sum_{i=1}^{q-1} a_i(Y_1)k^{i/d}N^{-i/d}+O\of{k^{q/d}N^{-q/d}}+O\of{1/k} & Y_1\in \mathcal{X_I^\kappa} \\
      O\of{(\kappa/N)^{q/d}} & Y_1\in \mathcal{X_B^\kappa},
   \end{cases}
\end{align}
where $a_i(Y_1)$ are constants depending on $Y_1$.

\end{lemma}
\begin{proof}
Suppose that the density $f$ is $q$ times differentiable, and all of the $q$ derivatives are bounded.
Let $y=Q_k(x)$. Also let $r=\rho_k(x)$, where $\rho_k(x)$ is defined as the $k$-NN distance on the point $x$. We can write $y=x+u\rho_k(x)$, where $u$ is unit vector. Then the Taylor expansion of $f(y)$ around $f(x)$ is as follows 
\begin{equation}\label{taylor}
f(y)=f(x)+\sum_{|i|\leq q}\frac{D^i f(x)}{i!}(u\rho_k(x))^i+O\of{\|u\rho_k(x)\|^q}.
\end{equation}

So we apply Lemma \ref{general_P_QinX_lemma} with the following choices for $\xi_j(Y_1)$, $j\in\{1,2\}$ 

\begin{align}
\xi_j(Y_1)&=\sum_{|i|\leq q}\frac{D^i f_j(Y_1)}{i!}(u\rho_k(Y_1))^i+O\of{\|u\rho_k(Y_1)\|^q},
\end{align}
which results in 

\begin{align}
\Pr\of{Q_k(Y_1)\in X}=\frac{f_1(Y_1)}{f_1(Y_1)+\eta f_2(Y_1)}+\tau_1(Y_1)+\tau_2(Y_1).
\end{align}

For the interior points, after simplifying $\tau_i(Y_1)$ given in equation \eqref{P_to_ratio_errors}, and using \eqref{KNN_bias} we get

\begin{align}
\tau_1(Y_1)+\tau_2(Y_1)=\sum_{i=1}^{q-1} a_i(Y_1)(k/N)^{i/d}+O\of{(k/N)^{q/d}}+O\of{1/k}, 
\end{align}
where $a_i(Y_1)$ is a constant depending only on $Y_1$.

For boundary points, by using a result in \cite{Poczos2014_1}(Bias Proof), we can bound the densities and get the desired upper bound. According to this result, for any $x\in \mathcal{X_B}$ and any $|i|<\gamma$, we have

\begin{align}\label{bound_derivative}
\abs{D^i f(x)}\leq \frac{Lh^{\gamma-|i|}}{(q-|i|)!},
\end{align}

where $h$ is the distance from $x$ to the boundary, and $L$ is a constant. Now note that since $\alpha_\kappa(x)<1$ and the $\kappa$-NN ball meets the boundary, we have $h<\rho_\kappa(x)$. Therefore, using the triangle inequality for \eqref{taylor} and setting $k=\kappa$, for every point $y=Q_\kappa(x)\in \mathcal{X}$ we have

\begin{align}
f(y) &\leq f(x)+\sum_{|i|\leq q} \abs{\frac{D^i f(x)}{i!}(u\rho_\kappa(x))^i }+O\of{\|u\rho_\kappa(x)\|^q}\nonumber\\
&= f(x)+\sum_{|i|\leq q} \abs{\frac{D^i f(x)}{i!}}\rho_\kappa^{|i|}(x)\abs{u^i }+O\of{\|u\rho_\kappa(x)\|^q}\nonumber\\
&\leq f(x)+\sum_{|i|\leq q} c_i\rho_\kappa^{q}(x)+O\of{\|u\rho_\kappa(x)\|^q} \nonumber\\
&=f(x)+O\of{\rho^q_\kappa(x)},
\end{align}
where in the third line we have used \eqref{bound_derivative} and the fact that $h<\rho_\kappa(x)$.
Using the bound on $k$-NN distances for the boundary points derived in Lemma \ref{KNN_bias_boundary_Lemma}, we have $\E{\rho^q_\kappa(x)}<O\of{(\kappa/N)^{q/d}}$.  
After simplifying $\tau_i(Y_1)$ given in equation \eqref{P_to_ratio_errors}, we get

\begin{align}
\tau_1(Y_1)+\tau_2(Y_1)= O\of{(\kappa/N)^{q/d}}.
\end{align}

The rest of the proof for both interior and boundary points follows similarly by replacing $O\of{\epsilon_{\gamma,k}}$ by $\tau_1(Y_1)+\tau_2(Y_1)$ in \eqref{point_probability_equation}, and finally we get a result similar to \eqref{Expec_ratio}.

\end{proof}

\begin{lemma}\label{bias_ensemble}
The bias of the estimator can be derived as follows
\begin{align}
\mathbb{B}\of[\widehat{J}_\alpha(X,Y)]= \sum_{i=1}^{q-1} \phi_{i,\kappa}(N) (k/N)^{i/d}+O\of{(k/N)^{q/d}}+O\of{1/k}.
\end{align}
\end{lemma}

\begin{proof}
Let define the notations $P_{\kappa,N}:=\Pr\of{Y_1\in \mathcal{X_I^\kappa}}$, $\bar{P}_{\kappa,N}:=1-P_{\kappa,N}$ and $\phi_{i,\kappa}(N):= P_{\kappa,N} \E{a_i(Y_1)\vert Y_1\in \mathcal{X_I^\kappa}}$. Using Lemma \ref{epsilon_ensemble} we have

\begin{align}
\mathbb{B}\of[\left(\frac{N_1}{M_1+1}\right)^\alpha] &= \Pr\of{Y_1\in \mathcal{X_I^\kappa}}\mathbb{B}\left[\left(\frac{N_1}{M_1+1}\right)^\alpha \middle\vert Y_1\in \mathcal{X_I^\kappa}\right] + \Pr\of{Y_1\in \mathcal{X_B^\kappa}}\mathbb{B}\left[\left(\frac{N_1}{M_1+1}\right)^\alpha\middle\vert Y_1\in \mathcal{X_B^\kappa}\right] \nonumber\\
&= \sum_{i=1}^{q-1} P_{\kappa,N}\E{a_i(Y_1)\vert Y_1\in \mathcal{X_I^\kappa}} (k/N)^{i/d}+P_{\kappa,N} O\of{(k/N)^{q/d}}+P_{\kappa,N} O\of{1/k}+ \bar{P}_{\kappa,N} O\of{(\kappa/N)^{\gamma/d}}\nonumber\\
&= \sum_{i=1}^{q-1} \phi_{i,\kappa}(N) (k/N)^{i/d}+ O\of{(k/N)^{q/d}}+O\of{1/k}.
\end{align}

Using equations \eqref{DtoJ_bias} and \eqref{J2sum} concludes the bias rate for $D_\alpha(X,Y)$.
\end{proof}

\begin{proof}[\textbf{Proof of Theorem \ref{ensemble_theorem}}] The proof follows by using the ensemble theorem in (\cite{Kevin16}, Theorem 4) with the parameters $\psi_i(l)=l^{i/d}$ and $\phi'_{i,d}(N)=\phi_{i,\kappa}(N)/N^{i/d}$.

\end{proof}

\section*{B. Variance Proof}
\begin{proof}[\textbf{Proof of Theorem \ref{variance}}]

First note that the variance proof for 
 $Y_i=\of{N_i/(M_i+1)}^{\alpha}$ and $\widehat{J}_\alpha(X,Y)$ is contained in the the proof for $\widehat{D}_\alpha(X,Y)$, and also the proof for $\widehat{D}_g(X,Y)$ is similar to that. So, here we only focus on the variance proof of $\widehat{D}_\alpha(X,Y)$.
 
Assume that we have two set of nodes $X_i$, $1\leq i \leq N$ and $Y_j$ for $1\leq j \leq M$. Without loss of generality, assume that $N < M$.  We consider the  $M-N$ virtual random points $X_{N+1},...,X_M$ with the same distribution as $X_i$, and define $Z_i:=(X_i,Y_i)$. Now for using the Efron\hyp Stein inequality on $Z:=(Z_1,...,Z_M)$, we consider another independent copy of $Z$ as  $Z':=(Z'_1,...,Z'_M)$ and define $Z^{(i)}:=(Z_1,...,Z_{i-1},Z'_i,Z_{i+1},...,Z_M)$. Let $\widehat{D}_\alpha(Z):=\widehat{D}_\alpha(X,Y)$ and $\widehat{J}_\alpha(Z):=\widehat{J}_\alpha(X,Y)$. Then, according to Efron\hyp Stein inequality we have

\begin{align} \label{ESTa_Var_proof_1}
\mathbb{V}\of[\widehat{D}_\alpha(Z)] &\leq \frac{1}{2} \sum_{i=1}^M \mathbb{E}\left[\left(\widehat{D}_\alpha(Z)-\widehat{D}_\alpha(Z^{(i)})\right)^2\right]. 
\end{align}
Using the Mean Value Theorem, and going back to the definition $\widehat{D}_\alpha(Z)=\frac{1}{\alpha-1}\log\of{\widehat{J}_\alpha(Z)}$, there exist some constant $C_{\alpha}$, such that

\begin{align}\label{DtoJ}
 &\half\sum_{i=1}^N \mathbb{E}\left[\left(\widehat{D}_\alpha(Z)-\widehat{D}_\alpha(Z^{(i)})\right)^2\right] \leq \half C_{\alpha} \sum_{i=1}^N \mathbb{E}\left[\left(\widehat{J}_\alpha(Z)-\widehat{J}_\alpha(Z^{(i)})\right)^2\right]. 
\end{align}
Therefore, we only need to bound the RHS of \eqref{DtoJ}, which is also an upper bound for $\mathbb{V}\of[\widehat{J}_\alpha(Z)]$.

\begin{align}\label{split}
\half \sum_{i=1}^N \mathbb{E}\left[\left(\widehat{J}_\alpha(Z)-\widehat{J}_\alpha(Z^{(i)})\right)^2\right]  &= \frac{1}{2} \sum_{i=1}^N \mathbb{E}\left[\left(\widehat{J}_\alpha(Z)-\widehat{J}_\alpha(Z^{(i)})\right)^2\right] \nonumber\\
&\qquad\qquad + \frac{1}{2} \sum_{i=N+1}^M \mathbb{E}\left[\left(\widehat{J}_\alpha(Z)-\widehat{J}_\alpha(Z^{(i)})\right)^2\right].
\end{align}

First, we give an upper bound on the first term in \eqref{split}, and the second term would be bounded similarly. Define

\begin{equation}\label{define_B}
 B_{\alpha,i} :=
\left(\frac{N_i}{M_i+1}\right)^\alpha-  \left(\frac{N^{(1)}_i}{M^{(1)}_i+1}\right)^\alpha
.
\end{equation}

Then we have

\begin{align}\label{variance_main}
\frac{1}{2} \sum_{i=1}^N \mathbb{E}\left[\left(\tilde{J}_\alpha(Z)-\tilde{J}_\alpha(Z^{(i)})\right)^2\right] &= \frac{N}{2} \mathbb{E}\left[\left(\tilde{J}_\alpha(Z)-\tilde{J}_\alpha(Z^{(1)})\right)^2\right]  \nonumber\\
&=\frac{N}{2}\E{\left(\frac{1}{M}\sum_{i=1}^M\left(\frac{N_i}{M_i+1}\right)^\alpha- \frac{1}{M}\sum_{i=1}^M\left(\frac{N^{(1)}_i}{M^{(1)}_i+1}\right)^\alpha \right)^2} \nonumber\\
&=\frac{N}{2M^2}\E{\left(\sum_{i=1}^M\left(\left(\frac{N_i}{M_i+1}\right)^\alpha-  \left(\frac{N^{(1)}_i}{M^{(1)}_i+1}\right)^\alpha \right)\right)^2} \nonumber\\
&=\frac{N}{2M^2}\E{\left(\sum_{i=1}^M B_{\alpha,i} \right)^2} \nonumber\\
&=\frac{N}{2M^2}\sum_{i=1}^M\E{ B_{\alpha,i}^2}+ \frac{N}{2M^2}\sum_{i\neq j}\E{ B_{\alpha,i}B_{\alpha,j}} \nonumber\\
&=\frac{N}{2M}\E{ B_{\alpha,2}^2}+ \frac{N}{2}\E{B_{\alpha,2}}^2,
\end{align}
where in the last line we used $\E{ B_{\alpha,i}B_{\alpha,j}}=\E{ B_{\alpha,i}}\E{B_{\alpha,j}}=\E{ B_{\alpha,i}}^2$ for $i\neq j$.
Next, we only need to find bounds on $\E{B_{\alpha,2}}$ and $\E{B_{\alpha,2}^2}$. In the following lemma we derive the essential bounds. 

\begin{lemma}\label{bound_B}
$\E{B_{\alpha,2}}$ and $\E{B_{\alpha,2}^2}$ satisfy the following relations:
\begin{equation}
\E{B_{\alpha,2}}=\E{B_{\alpha,2}^2}=O(\frac{1}{N}).
\end{equation}
\end{lemma}

\end{proof}

\begin{proof}
The proof is similar for $\E{B_{\alpha,2}}$ and $\E{B^2_{\alpha,2}}$. So here we only focus on $\E{B_{\alpha,2}}$. We can assume that we re-sample $X$ and $Y$ separately, and both of the events are similar. Let $B^X_{\alpha,2}$ and $B^Y_{\alpha,2}$ denote the re-sampling difference in \eqref{define_B} when we only re-sample either $X_1$ or $Y_1$ points, respectively. Then it is easy to show that $\E{B_{\alpha,2}}\leq\E{ B^X_{\alpha,2}}+\E{B^Y_{\alpha,2}}$. 

Considering the re-sampling of $X_1$, we can write

\begin{align}\label{E_B_X1}
\E{B^X_{\alpha,2}}=\sum_{i=0}^1\sum_{j=0}^1\Pr(E_{ij})\E{B_{\alpha,2}|E_{ij}},
\end{align}

where $E_{00}$ is the event that none of $X_1$ and $X'_1$ fall within $k$ nearest neighbor points of $X_2$, $E_{01}$ is the event that $X_1$ and $X'_1$ fall within and not among the $k$ nearest neighbor points of $X_2$, respectively, $E_{10}$ is the event that $X'_1$ and $X_1$ fall within and not among the $k$ nearest neighbor points of $X_2$, respectively, and finally $E_{11}$ is the event that both of $X_1$ and $X'_1$ fall within $k$ nearest neighbor points of $X_2$. Now Note that in both of the events of $E_{00}$ and $E_{11}$ we have $B_{\alpha,2}=0$. Also since the events $E_{01}$ and $E_{10}$ are symmetric, we only consider the event $E_{01}$:

\begin{align}
\Pr(E_{01})=(\frac{k}{N})(1-\frac{k}{N})= O(\frac{k}{N}).
\end{align}

Going back to \eqref{E_B_X1}, we have

\begin{align}\label{E_B}
\E{B^X_{\alpha,2}}&=O(\frac{k}{N})\E{B^X_{\alpha,2}|E_{01}}\nonumber\\
&=O(\frac{k}{N})\left(\E{\frac{N_2^\alpha-(N_2+1)^{\alpha}}{(M_2+1)^\alpha}}\frac{f_1(X_2)}{f_1(X_2)+\eta f_2(X_2)}+\E{\frac{N_2^\alpha}{(M_2+1)^\alpha}-\frac{N_2^\alpha}{(M_2+2)^\alpha}}\frac{\eta f_2(X_2)}{f_1(X_2)+\eta f_2(X_2)}\right).
\end{align}

By using Taylor expansion, there exist a constant $e_1$ such that

\begin{align}
\E{\frac{N_2^\alpha-(N_2+1)^{\alpha}}{(M_2+1)^\alpha}}\leq\E{(M_2+1)^{-1}\frac{e_1N_2^{\alpha-1}}{(M_2+1)^{\alpha-1}}}.
\end{align}

Note that $N_2^{\alpha-1}/(M_2+1)^{\alpha-1}$ is bounded from above by $(C_U/C_L)^{\alpha-1}$. Also from \eqref{E_M} we get $\E{(M_2+1)^{-1}}=O\of{1/k}$. Thus,

\begin{align}
\E{\frac{N_2^\alpha-(N_2+1)^{\alpha}}{(M_2+1)^\alpha}}\frac{f_1(X_2)}{f_1(X_2)+\eta f_2(X_2)} \leq O\of{1/k}.
\end{align}

We can similarly show that 

\begin{align}
\E{\frac{N_2^\alpha}{(M_2+1)^\alpha}-\frac{N_2^\alpha}{(M_2+2)^\alpha}}\frac{f_2(X_2)}{f_1(X_2)+\eta f_2(X_2)}\leq O(1/k).
\end{align}

So, as result \eqref{E_B} becomes

\begin{align}
\E{B^X_{\alpha,2}}=O(1/N).
\end{align}

Using a similar approach one can simply show that $\E{B^Y_{\alpha,2}}=O(1/N)$.
So, finally we have $\E{B_{\alpha,2}}=O(1/N)$.
\end{proof}

From \eqref{variance_main} and Lemma \ref{bound_B} we get 

\begin{align}
\frac{1}{2} \sum_{i=1}^N \mathbb{E}\left[\left(\tilde{J}_\alpha(Z)-\tilde{J}_\alpha(Z^{(i)})\right)^2\right] \leq O\of{1/M}+O\of{1/N}.
\end{align}

Using a similar approach, we can also simply show that 

\begin{align}
\sum_{i=N+1}^M \mathbb{E}\left[\left(\widehat{J}_\alpha(Z)-\widehat{J}_\alpha(Z^{(i)})\right)^2\right]\leq O\of{1/M}+O\of{1/N}.
\end{align}

Finally using \eqref{ESTa_Var_proof_1}, \eqref{DtoJ} and \eqref{split} we get

\begin{align}
\mathbb{V}\of[\widehat{D}_\alpha(Z)] &\leq  O\of{1/M}+O\of{1/N}.
\end{align}


\end{document}